\documentclass[12pt]{iopart}
\usepackage{iopams}  
\usepackage{graphicx}
\begin{document}

\title[Lepton asymmetry and the cosmic QCD transition]{Lepton asymmetry and the cosmic QCD transition}

\author{Dominik J Schwarz and Maik Stuke}

\address{Fakult\"at f\"ur Physik, Universit\"at Bielefeld, Postfach 100131, 33501 Bielefeld, Germany}
\eads{\mailto{dschwarz} and \mailto{mstuke} at \mailto{physik.uni-bielefeld.de}}

\begin{abstract}
We study the influence of lepton asymmetries on the evolution of the early Universe. The lepton asymmetry $l$ is poorly constrained by observations and might be orders of magnitudes larger 
than the observed baryon asymmetry $b \simeq 10^{-10}$, $|l|/b \leq 2\times 10^8$. We find that lepton asymmetries that 
are large compared to the tiny baryon asymmetry, can influence the dynamics of the QCD phase 
transition significantly. The cosmic trajectory in the $\mu_B-T$ phase diagram of strongly interacting matter becomes a function of lepton (flavour) asymmetry. For tiny or vanishing baryon and lepton asymmetries lattice QCD simulations show that the cosmic QCD transition is a rapid crossover. However, for large lepton asymmetry, the order of the cosmic transition remains unknown.
\end{abstract} 

\pacs{98.80Es, 12.38Aw}
\vspace{2pc}
\noindent{\it Keywords}: early Universe, cosmic QCD transition, lepton asymmetry
%\maketitle
%%%%%%%%%%%%%%%%%%%%%%%%%%%%%%%%%%%%%%%%%%%%%%%%%%%%%%%%%%%%%%%%%%%%%%%%

\section{Introduction}
One of the most poorly constrained parameters of cosmology is the specific lepton asymmetry $l$,
the lepton number per entropy, mainly because the cosmic neutrino background escapes all 
direct attempts to be measured. The strongest constrains stem from a combined analysis 
of big bang nucleosynthesis (BBN) and the cosmic microwave background (CMB), leading to 
$|l| \leq 0.02$ \cite{Serpico:2005bc, Simha:2008mt, Popa:2008tb}. 

Cosmological inflation predicts for the observable universe that both, baryon and lepton numbers are approximately zero. Therefore baryo- or leptogenesis or a combination of both is needed to result in the measured specific baryon asymmetry $b \simeq 10^{-10}$. 
Before the electroweak transition at $T_{\rm EW} \sim 200$ GeV, 
sphaleron processes allow the violation of baryon and lepton number in a way, that 
$\Delta(B - L)=0$. 
Within the standard model of particle physics (SM) this would lead to $l = -(51/28)b$,
where the numerical coefficient depends on the particle content of the universe before the onset of the electroweak transition 
\cite{Harvey:1990qw}. However, the asymmetries might differ by orders of magnitudes.  Late (compared to the epoch when sphaleron processes cease to be efficient) baryogenesis would lead to 
$b \gg | l |$, while late leptogenesis would lead to $|l|\gg b$ 
\cite{D'Ambrosio:2003wy, D'Ambrosio:2004fz, Covi:1996wh, Flanz:1996fb,Hambye:2001eu}. 
From the experimental point of view, such a large specific lepton asymmetry may be hidden in 
todays neutrino sector. 

A likewise unknown, although most dramatic event in the early universe is the quark-hadron transition at approximately $10 \mu$s after the big bang, where quarks and gluons condense to hadrons.
The SM predicts a spontaneous breaking of the chiral 
symmetry of QCD and a confining of quarks into hadrons at a (pseudo-)critical Temperature $T_c$. 
This epoch is one of the most interesting in the early universe, but experimentally only little is known about this QCD transition at $T_c \approx 190$ MeV \cite{Cheng:2006qk, Karsch:2007zza}. 
It is well established that the order of the transition depends on the baryon density, or equivalently 
the baryochemical potential $\mu_B$. It is common to draw a QCD phase diagram in the $\mu_B-T$ plane, in which the cosmic QCD transition is commonly assumed to take place at $T=T_c$ and $\mu_B \approx 0$. We show in 
this work that this assumption holds only if $|l| = {\cal O}(b)$. For $|l|\gg b$, the transition happens at $\mu_B={\cal O}(l)T_c$.

The QCD transition is in the focus of the relativistic heavy-ion research programs at RHIC and LHC. 
These experiments take place at very small baryochemical potential and high temperatures, presumably far away from a critical point. More information about a possible first-order transition 
at large values of $\mu_B$ are expected from the future FAIR program. Todays ordinary matter is located at $T\approx0$ and $\mu_B$ equal to the nucleon mass. Extremely dense matter, like the interior of compact stars with $T \leq 10$ MeV, are expected to reach some colour superconducting phase \cite{Alford:2007xm}. 

\begin{figure}
        \centering
        \includegraphics[angle=270,width=0.8\textwidth]{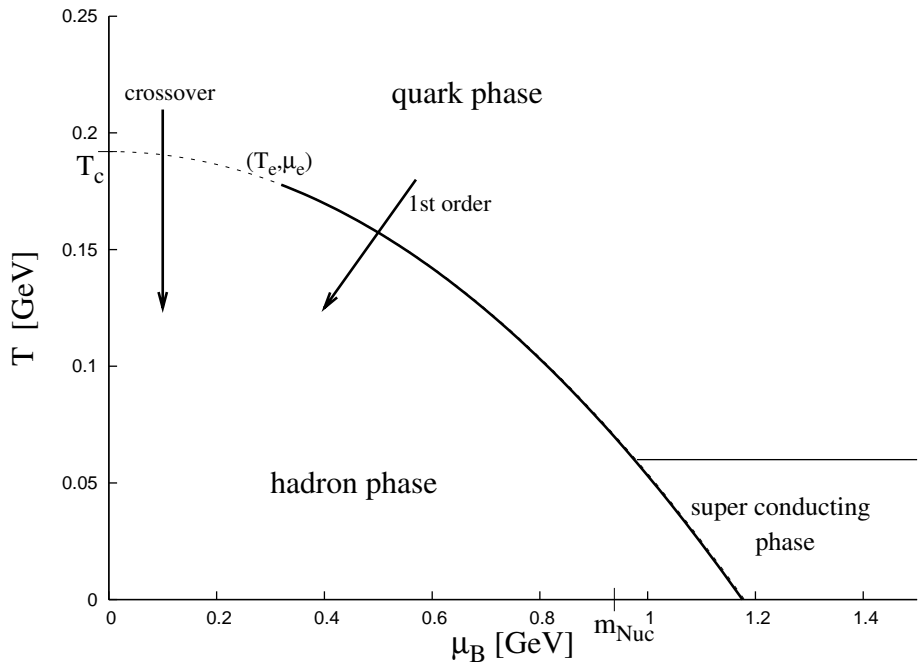}
        \caption{Sketch of the phase diagram for physical QCD, 
         based on findings from nuclear physics, lattice QCD and perturbative calculations. 
         Solid lines indicate a first-order phase transition, while the dashed line indicates a crossover. 
         The exact phase diagram with or without a critical end point ($T_{\rm e},\mu_{\rm e}$) 
         is still under debate. So is also the (pseudo-)critical temperature
         $T_{\rm c} = 192(7)(4)$ MeV \cite{Cheng:2006qk,Karsch:2007zza},  which differs from  
         $T_{\rm c} = 164\pm 2$ MeV, the value found in \cite{Aoki:2006br}. 
         While \cite{de Forcrand:2007zz} argue that it is still an
         open question if a critical end point exists at all, \cite{Fodor:2004nz} find 
         $\mu_{\rm e} = 360 \pm 40$ MeV and $T_{\rm e} = 162 \pm 2$ MeV. 
         The calculations and methods leading to these different results are discussed and 
         compared in \cite{Karsch:2007zza,Aoki:2006br,de Forcrand:2007zz,Stephanov:2007fk}.}
         \label{fig:phdia2}
\end{figure}

For $T$ and $\mu_B$ being of the order of the QCD confining scale $\Lambda_{\rm QCD}$, the most reliable tool is lattice simulations with quarks. As a tribute to different possible solutions to the so called `sign problem' \cite{Stephanov:2007fk, Schmidt:2007jg}, different phase diagrams with different $T_c$ and different critical lines have been proposed 
\cite{Philipsen:2007rm, Stephanov:2007fk, de Forcrand:2007zz}. 
The widely expected shape of the QCD phase diagram in a $\mu_B-T$ plane as a combination of results from nuclear physics, perturbative calculations and lattice simulations is shown in figure 
\ref{fig:phdia2}.

The cosmic QCD transition sets the initial condition for big bang nucleosynthesis (BBN). 
Standard BBN relies on the assumption that the Universe can be treated as a homogeneous and isentropic radiation fluid at the moment of BBN. Therefore it is of great interest to know exactly what happens at the QCD transition \cite{Applegate:1985qt}. The cosmic QCD transition might also create some observable relics. Most of them can only be generated during a first-order transition, 
like quark nuggets or magnetic fields \cite{Witten:1984rs, Schwarz:2003du}, while QCD balls could 
be generated independently of the order of the transition \cite{Zhitnitsky:2002nr}. The transition 
might also lead to the formation of cold dark matter (CDM) clumps \cite{Schmid:1996qd} and definitely leads to a modification of the primordial background of gravitational waves \cite{Schwarz:1997gv, Watanabe:2006qe}. 

In this work, we trace the baryo-, lepto- and charge chemical potentials in the early Universe as a 
function of temperature around the QCD phase transition. We assume that the Universe is neutral and that electric charge, baryon and lepton (flavor) numbers are conserved. For vanishing lepton 
asymmetry this situation has been studied in \cite{Rafelski:2007zz}. The effect of lepton asymmetries, but neglecting particle masses, was first considered in \cite{Zarembo:2000wj}. 

In section 2 we explain our assumptions about the particle content and their interactions in the early universe after the electroweak transition and before BBN. We introduce the relevant conserved quantum numbers and use the corresponding conservation laws to solve for the net particle densities and corresponding particle chemical potentials in the quark and hadron phases. 
In the third section, we introduce the chemical potentials related to the conserved quantum numbers 
and study the trajectories of the Universe in the corresponding $\mu-T$ planes, depending on the assumed lepton asymmetry. We discus some consequences of our findings on the cosmic QCD transition in section 4 and conclude in section 5.

\section{Evolution of particle chemical potentials and densities}

In contrast to the treatment of the QCD transition in colliders, where non-equilibrium effects play an important role, for the cosmic transition thermal and chemical equilibrium are excellent approximations.
For the thermodynamic description of the early universe, the Hubble time, $t_{\rm{H}}=1/\rm{H}$, is the timescale of interest \cite{Schwarz:2003du}. At $T_{\rm{QCD}}\simeq 200$ MeV it is $t_H\simeq 10^{-5}$ s. This is a long time, compared to the timescales of strong, electromagnetic, and even weak interactions with $1/G_{\rm{F}}^2T_{\rm{QCD}}^5\simeq 10^{-9}$ s. 

In the early Universe, after the electroweak phase transition at $T_{\rm EW} \sim 200$ GeV and 
before the onset of neutrino oscillations at a few MeV, baryon number $B$, lepton flavour numbers 
$L_f$, and the 
electric charge $Q$ are conserved. Additionally, we assume that $Q=0$. A neutral universe seems 
to be reasonable, because several observations point to neutrality and lack of currents on large 
scales. The work of \cite{Siegel:2006px} shows that a possible charge asymmetry is annihilated in less than a Hubble time for $100$ GeV $\geq T \geq 1$ eV. It is also shown, that currents are damped for all temperatures $T\geq 1$ eV.

We assume that all globally conserved quantum numbers are conserved locally, too. 
This means that our approximation is applicable at length scales larger than the largest scale on which transport phenomena can show up, given we apply homogeneous initial conditions. The largest mean free path is that of neutrinos and thus the local physics on scales less than the neutrino mean free 
path might differ from the results obtained in this work. 

The net particle density $n_i$ is the density of a particle minus the density of its anti-particle. In 
thermal and chemical equilibrium (and neglecting effects of interactions) it can be described as
\begin{equation}
        \fl     n_i (T, \mu_i) = \frac{g_i}{2\pi^2}\int_{m_i}^{\infty}{E\sqrt{E^2-{m_i}^2}}
        \left(\frac{1}{\rm{exp}(\frac{E-\mu_i}{T})\pm 1}-\frac{1}{\rm{exp}(\frac{E+\mu_i}{T})\pm 1}\right)dE, 
\label{nifull}
\end{equation} 
for a particle with mass $m_i$, chemical potential $\mu_i$ and helicity degrees of freedom $g_i$. The 
sign is fixed by the spin of the particle, $+$ for fermions, $-$ for bosons. As the Universe expands and cools, the particles annihilate at temperatures comparable to their mass and eventually fall out 
of equilibrium.

Lifting the global conservation laws to local ones we obtain equations for the specific lepton flavour asymmetry, specific baryon asymmetry and the charge density:
\numparts
\begin{eqnarray}
\fl   l_f  = & \frac{n_f + n_{\nu_f}}s  \quad & \mbox{for\ }f=e,\mu,\tau, \\
\fl   b    = & \sum_{i}\frac{b_i n_i}s \quad & \mbox{with\ }b_i = \rm{baryon\ number\ of\ species\ }i, \\
\fl   0    = & \sum_{i}{q_i n_i}           \quad & \mbox{with\ }q_i = \rm{charge\ of\ species\ }i,
\end{eqnarray}
\endnumparts
with the entropy density $s=s(T)$. For any given temperature $T$, the free parameters in this set of equations are the chemical potentials in the net particle densities $n_i(T,\mu_i)$ and the 
specific lepton flavour asymmetries $l_f$. The baryon number $b \approx 
 (8.85\pm 0.24) \times 10^{-11}$ is fixed by observations \cite{Komatsu:2008hk}.

In the early Universe particle densities are usually not high enough for many body reactions. The main processes that keep all particles in thermal and chemical equilibrium are $2\leftrightarrow2$-processes. The chemical potentials of photons and gluons are zero, $\mu_{\gamma}=\mu_g=0$, as they  can be produced and annihilated in any number. Consequently, the chemical potentials of a particle $i$ and its anti-particle $\bar{i}$ are equal, but of opposite sign: $\mu_i=-\mu_{\bar i}$.

The leptons are then linked via reactions like $f_i + \bar{f}_j \rightleftharpoons 
\nu_{f_i} + \bar{\nu}_{f_j}$ for all combinations of flavours, 
leading to relations like 
\begin{equation}
\label{muemumu}
\mu_{e} - \mu_{\nu_e} = \mu_{\mu} - \mu_{\nu_{\mu}}.
\end{equation}

\subsection{Quark phase}

Due to flavour mixing in the quark sector, we only need to distinguish between up and down 
type quarks with chemical potentials $\mu_u = \mu_c =\mu_t$ for the up, charm and top quarks and 
$\mu_d=\mu_s=\mu_b$ for the down, strange and bottom quarks. These quark chemical potentials are further linked via weak reactions like $u+e^-\rightleftharpoons d+\nu_e$, 
which leads to $\mu_d=\mu_u+\mu_e-\mu_{\nu_e}$. 

The number of independent chemical potentials in the quark phase is thus reduced to four leptonic potentials 
($\mu_e$, $\mu_{\nu_e}$, $\mu_{\nu_{\mu}}$, $\mu_{\nu_{\tau}}$) and just one for the quarks 
($\mu_u$).  For temperatures well below $m_W$ 
and in chemical and thermal equilibrium, the evolution of the 
net particle densities as a function of temperature is described by five conservation equations
\numparts
\begin{eqnarray}
\label{ne}
	0 = & n_e(T, \mu_e) + n_{\nu_e}(T,\mu_{\nu_e}) - l_e s(T), \\
\label{nmu}
	0 = & n_{\mu}(T, \mu_{\nu_{\mu}},\mu_e,\mu_{\nu_e}) + n_{\nu_{\mu}}(T, \mu_{\nu{\mu}})
	- l_{\mu} s(T), \\
\label{ntau}
	0 = & n_{\tau}(T, \mu_e,\mu_{\nu_e},\mu_{\nu_{\tau}}) + n_{\nu_{\tau}}(T, \mu_{\nu_{\tau}}) 
	- l_{\tau} s(T), \\
\label{B=0}				
	0 = & n_u(T, \mu_u) + n_d(T, \mu_u,\mu_e,\mu_{\nu_e}) + n_c(T, \mu_u) + 
	          n_s(T, \mu_u,\mu_e,\mu_{\nu_e}) \\ 
\nonumber
	       & + n_b(T, \mu_u,\mu_e,\mu_{\nu_e}) - 3b s(T), \\
\label{q0end} 
	0 = & n_u(T, \mu_u) + n_c(T, \mu_u) - n_e(T, \mu_e) 
	- n_{\mu}(T, \mu_{\nu_{\mu}},\mu_e,\mu_{\nu_e}) \\ 
\nonumber 
	       & - n_{\tau}(T, \mu_e,\mu_{\nu_e},\mu_{\nu_{\tau}}) - b s(T).
\end{eqnarray}   
\endnumparts
The last equation combines the conservation of electric charge and baryon number in a convenient manner.  Exact solutions to this system of coupled integral equations must be obtained numerically. However,  before doing so, we estimate the evolution of the chemical potentials at high temperatures
via an analytic approach. 

At $T > T_c$, but close to the QCD transition, we consider the three light quarks $u$, $d$ and $s$, 
as well as electrons and muons with their corresponding neutrinos (the masses of these particles are neglected). Tau leptons annihilate well before the QCD transition (at $T_\tau \sim 600$ MeV), 
hence tau neutrinos are decoupled at the temperatures of interest. For relativistic particles 
($T\gg m_i,\mu_i$) the net number densities become
\begin{eqnarray}        
        \label{Tbig} n_i & = \frac{1}{3} g_iT^2\mu_i +O(\mu_i^3)\ \ \ \rm{for\ bosons}, \\ 
                                     &=\frac{1}{6} g_iT^2\mu_i +O(\mu_i^3)\ \ \ \rm{for\ fermions},
\end{eqnarray}
while we neglect the non-relativistic particles. With these approximations, 
equations (\ref{ne}) to (\ref{ntau}) become:
\numparts
\begin{eqnarray}
      \label{leAna} 
      l_e s(T) & = \frac 16 \left(2\mu_e+\mu_{\nu_e} \right) T^2, \\
      \label{lmuAna}
      l_{\mu} s(T) & = \frac 16 \left(2\mu_{\mu}+\mu_{\nu_{\mu}} \right) T^2,\\
      \label{ltauAna}
      l_{\tau} s(T) & = \frac 16 \mu_{\nu_{\tau}} T^2.
\end{eqnarray}
\endnumparts
Let us further assume (for simplicity) that $l_e = l_\mu = l_\tau = l/3$, which implies, 
together with (\ref{muemumu}), that electrons and muons, as well as electron 
and muon neutrinos have equal chemical potentials, while (\ref{ltauAna}) fixes  
$\mu_{\nu_\tau} = 2l s(T)/T^2$. 
Consequently, from the complete set of conservation equations, the chemical potentials for 
all particles can be expressed in terms of the lepton and baryon asymmetries, $l$ and $b$
(for three quark and two lepton flavours):
\numparts
\begin{eqnarray}
\label{muusimp}        \frac{\mu_u}{T} &=& \frac{b s(T)}{T^3} + \frac{1}{3}\frac{l s(T)}{T^3},\\
\label{mudsimp}        \frac{\mu_d}{T}&=& \frac{b s(T)}{T^3} - \frac{1}{6}\frac{l s(T)}{T^3},\\
\label{muesimp}				\frac{\mu_e}{T} &=& \frac 12 \frac{\mu_{\nu_e}}{T} = \frac{1}{2}\frac{l s(T)}{T^3}.
\end{eqnarray}
\endnumparts
While the chemical potentials for the leptons (\ref{muesimp}) depend on the lepton asymmetry only, 
the quark chemical potentials given by (\ref{muusimp}) and (\ref{mudsimp}) depend on baryon and lepton asymmetry. 

We would like to stress that the result (\ref{muusimp}) to (\ref{muesimp}) strongly depends on the 
number of relativistic quark and lepton flavours. E.g. increasing the temperature to $T \sim 1$ GeV, 
the inclusion of tau leptons and charm quarks modifies the coefficients in the above result, i.e. 
\numparts
\begin{eqnarray}
\label{muu1GeV}        \frac{\mu_u}{T} &=& \frac{9}{16} \frac{b s(T)}{T^3} + \frac{1}{4}\frac{l s(T)}{T^3}, \\
\label{mud1GeV}        \frac{\mu_d}{T}&=&\frac{15}{16} \frac{b s(T)}{T^3} - \frac{1}{4}\frac{l s(T)}{T^3}, \\
\label{mue1GeV}	  \frac{\mu_e}{T} &=& \frac{1}{2}\frac{l s(T)}{T^3} + \frac{1}{8} \frac{b s(T)}{T^3}, \\
\label{munu1GeV}      \frac{\mu_{\nu_e}}{T} &=& \frac{l s(T)}{T^3} - \frac{1}{4} \frac{b s(T)}{T^3}.
\end{eqnarray}
\endnumparts
It actually turns out that the previously found independence of the chemical potentials of the 
leptons from the baryon asymmetry is a coincidence of the three quark flavour case. The deeper reason 
is that the charge of one down and one strange quark just compensate the charges of one up quark.
 
In general, \textit{all} particle chemical potentials depend on $b$ \textit{and} $l$. For three quark flavours this is 
also the case when the strange quark mass is taken into account.
For physical particle masses (e.g.~the strange quark mass and the mass of the muon 
cannot be neglected during the QCD transition) an analytic solution is not possible. The full 
numerical solution is presented below.

\subsection{Hadron phase}

The hadron phase (for $T <  T_c$) contains besides hadrons, electrons and muons as well as 
all three types of neutrinos. At low temperatures, all quarks are bound in neutrons and protons, but close to the QCD transition also mesons, mainly pions, are produced in significant numbers. 
The conservation of quantum numbers, expressed in terms of the net particle densities depending on four independent chemical potentials, gives rise to the relations:
\numparts
\begin{eqnarray}
\label{LeQCD}   
	0 & = n_e(T,\mu_e)-n_{\nu_e}(T,\mu_{\nu_e}) - l_e s(T), \\
\label{LmuQCD} 
	0 & = n_{\mu}(T, \mu_p,\mu_n,\mu_{\nu_{\mu}}) - n_{\nu_{\mu}}(T, \mu_{\nu_{\mu}}) - l_{\mu} s(T), \\
\label{BQCD}    
	0 & = n_p(T, \mu_p)+n_n(T, \mu_n)-b  s(T), \\
\label{Q0QCD}   
	0 & = n_p(T, \mu_p)-n_e(T, \mu_e)-n_{\mu}(T, \mu_p,\mu_n,\mu_{\nu_{\mu}})-n_{\pi}(T, \mu_p,\mu_n).
\end{eqnarray}
\endnumparts 
For a simplified analytic treatment, we neglect all mesons and muons.  Only protons, neutrons, 
electrons and electron neutrinos are taken into account. The leptons are still relativistic (as long as 
$T > m_e/3$), while protons and neutrons are non-relativistic. Their net particle density is approximated 
as 
\begin{equation}
n_i \simeq 2 g_i \left(\frac{m_iT}{2\pi} \right)^{3/2} 
\mathrm{sinh}\left(\frac{\mu_i}{T}\right)\mathrm{exp}\left(\frac{-m_i}{T} \right).
\end{equation}
The system of equations (\ref{LeQCD}), (\ref{BQCD}), and (\ref{Q0QCD}) then becomes (as above, we
assume $l_e = l/3$)
\numparts
\begin{eqnarray}
        \label{AnaQCDLe} l s(T) & = \frac12 T^2(2\mu_e+\mu_{\nu_e}),\\
        \label{AnaQCDB} b s(T) & =
                        4\left(\frac{Tm}{2\pi}\right)^{3/2}\mathrm{exp}\left
                        (\frac{-m}{T}\right)\left(\mathrm{sinh}\frac{\mu_p}{T}+
                        \mathrm{sinh}\frac{\mu_n}{T} \right),\\
        \label{AnaQ0QCD} 0 & =
                        4\left(\frac{Tm}{2\pi}\right)^{3/2}\mathrm{exp}\left
                        (\frac{-m}{T}\right)\mathrm{sinh}\frac{\mu_p}{T}-\frac{1}{3}T^2\mu_e, 
\end{eqnarray} 
\endnumparts
where we assumed $m_n \approx m_p \equiv m$. The weak reaction 
$e+p \rightleftharpoons n+\nu_e$, which are in equilibrium at energies above $\sim 1$ MeV, (\ref{AnaQCDLe}) imply
\begin{equation}
\label{mueHana}
\frac{\mu_e}{T} = \frac{2l s(T)}{3T^3} + \frac{\mu_n-\mu_p}{3T}.
\end{equation}
For $\mu_p/T \ll 1$ and $\mu_n/T \ll 1$, we may simplify (\ref{AnaQCDB}) to obtain
\begin{equation}
\label{mupmun}
\frac{\mu_p}{T}+\frac{\mu_n}{T} = \frac{bs(T)}{c(T)} \exp[m/T],
\end{equation}
with $c(T) = 4 (\frac{Tm}{2\pi})^{3/2}$. Together with (\ref{mueHana}) and (\ref{AnaQ0QCD})
the proton chemical potential reads
\begin{equation}
\label{mupHana}
\frac{\mu_p}{T} = 
\frac{\frac{l s(T)}{ T^3} + \frac{b s(T)}{2 c(T)} \exp[m/T]}{1+\frac{9 c(T)}{2T^3}\exp[-m/T]}.
\end{equation}
Note that this expression holds as long as $\mu_p/T \ll 1$. Nevertheless, the expression 
allows us to understand why the lepton asymmetry does not couple to the proton and 
neutron chemical potentials. As $T$ decreases, the influence of the baryon asymmetry eventually 
overcomes any lepton asymmetry, as the term $\propto b \exp[m/T]$ grows exponentially. 
Thus, in the low temperature regime, large lepton asymmetries play no role for the 
proton chemical potential. In the same limit, $\mu_p \approx \mu_n$. 
From equation (\ref{mueHana}) we find that a large lepton asymmetry results in large neutrino and electron asymmetries, as $\mu_{\nu_e} \approx \mu_e$.

As the temperature drops further, $\mu_p/T$ becomes large and the above approximation breaks 
down.  In that case we can approximate $\mathrm{sinh}(\mu/T)$ in (\ref{AnaQCDB}) by 
$\mathrm{exp}(\mu/T)/2$ and we finally find that 
\begin{equation}
\label{lowTmup}
\mu_p \approx m - T \ln [c(T)/2 b s(T)].
\end{equation}
In the low temperature regime, $\mu_p$ is linear in $T$ and for small $T$ it runs against the mass $m$. The electron asymmetry finally is just the same as the proton asymmetry and a possible large lepton asymmetry is turning into a large neutrino asymmetry
in the late Universe.   

Thus, large lepton asymmetries in  the low temperature regime of the hadron gas affect only the chemical potentials of leptons and play no role for $\mu_p$ and $\mu_n$.

\subsection{Numerical results}

Taking all particle masses into account, equations (\ref{ne}) to (\ref{q0end}) and 
(\ref{LeQCD}) to (\ref{Q0QCD}) have to been solved numerically. We use Gauss-Laguerre 
integrations for the net particle densities and a multidimensional secant method, 
called \textit{Broydn's method} \cite{Press:1996pp}.
To avoid numerical instabilities, the net particle densities in (\ref{nifull}) are rewritten as
\begin{equation}
n=\frac{g}{2\pi}\ \rm{sinh}\left[\frac{\mu}{T} \right]\rm{exp}\left[ -\frac{m}{T}\right]{(Tm)}^{3/2}I(T,\mu), \nonumber
\end{equation}
where
\begin{equation}
I(T,\mu)=\int^{\infty}_0\frac{(1+\frac{T}{m}x)\sqrt{(1+\frac{T}{m}x)x}}{\left(\rm{exp}[x-\frac{\mu}{T}]+\rm{exp}[-\frac{m}{T}] \right)\left(\rm{exp}[x+\frac{\mu}{T}]+\rm{exp}[-\frac{m}{T}] \right)}{\rm d}x. \nonumber
\end{equation}

For the quark phase we consider the temperature interval $10$ GeV  $>T> 10$ MeV with all SM 
particles, even if some of them, like the top-quark or the $W^{\pm}$ and Z-bosons are to heavy to play a significant role in this temperature regime. All particle masses are adopted from the Particle Data Group 
\cite{Amsler:2008zzb}. We restrict our study to the case of equal flavour asymmetries 
$l_{e}=l_{\mu}=l_{\tau}=l/3$. The baryon asymmetry is fixed to $b = 9 \times 10^{-11}$ in all our 
numerical calculations.  As already discussed above, if sphaleron processes are efficient and no additional baryon or leptogenesis happens after they stop, the generic value of 
$l  = -\frac{51}{28}b$ \cite{Harvey:1990qw}. This leads to a universe dominated by antimatter in the leptonic sector and matter 
in the baryonic sector. Below we discuss two examples in detail, $l = -b$ and $l = 3 \times 10^{-4}$.

\begin{figure}
        	\centering
        	\includegraphics[angle=270,width=0.8\textwidth]{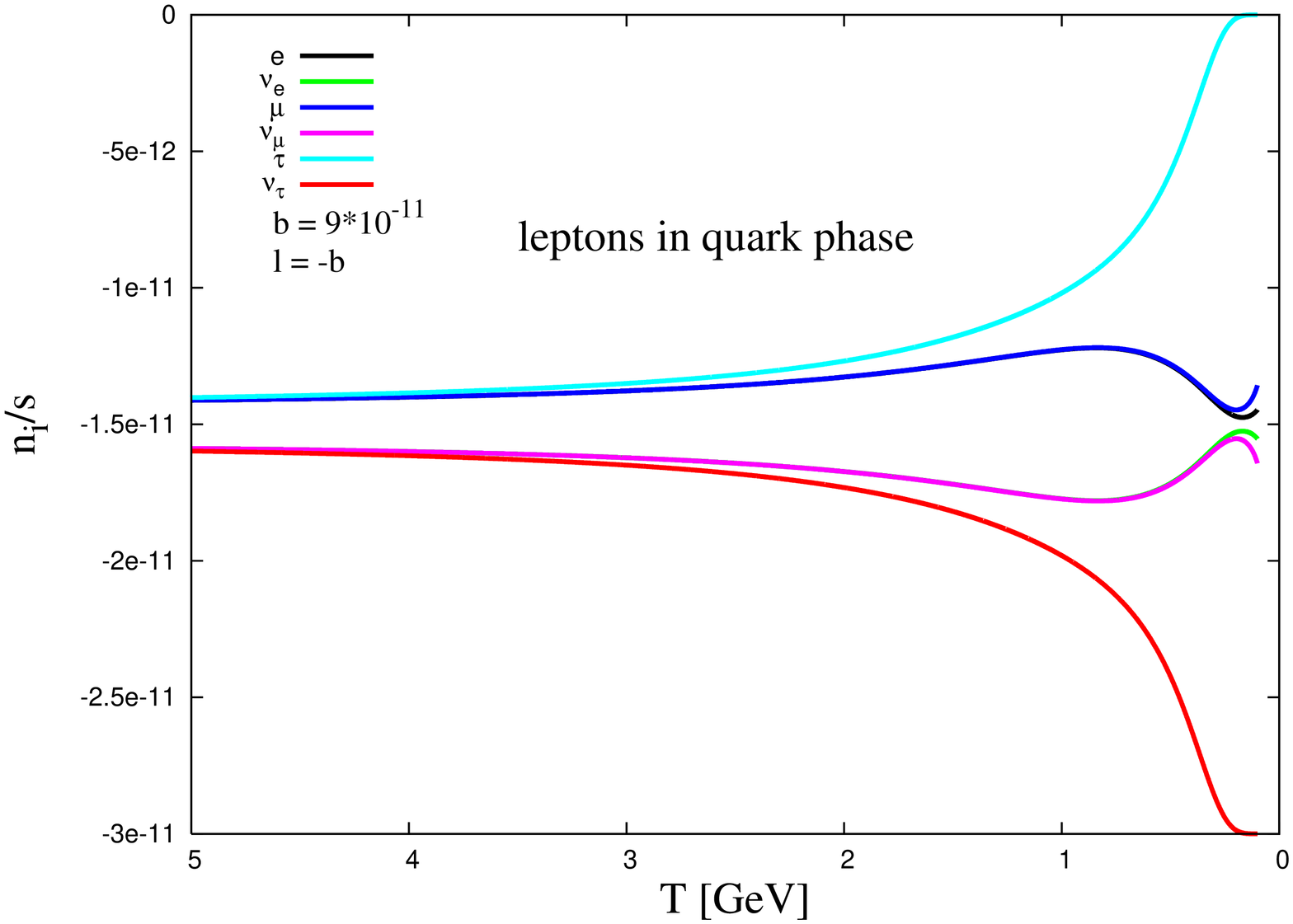}
        	\caption{Evolution of net lepton densities $n_i$ in the quark phase for $l=-b$. We plot them with respect to the entropy density $s(T).$}
        	\label{fig:figure2}
        	\vspace{1cm}
        	\includegraphics[angle=270,width=0.8\textwidth]{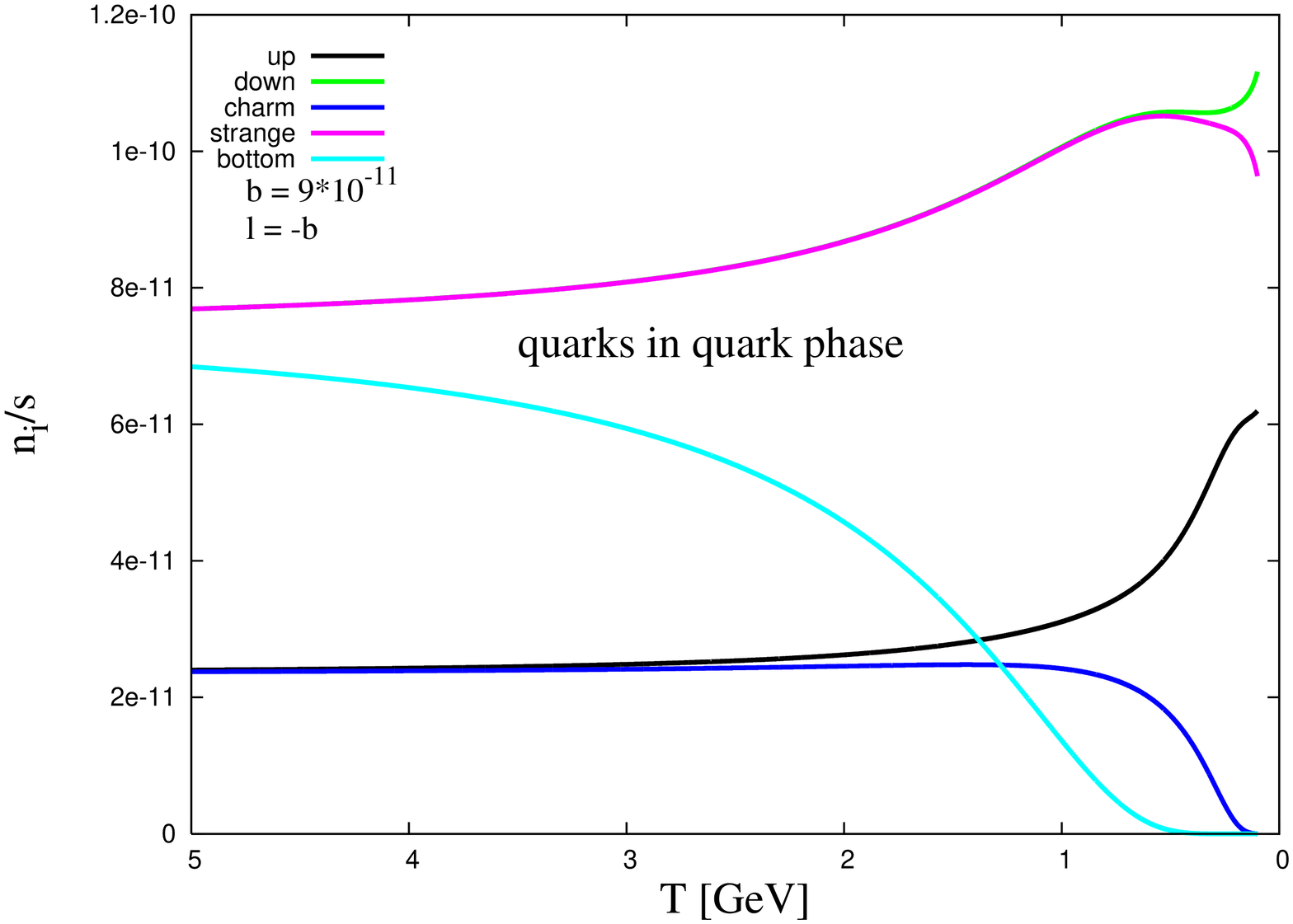}
        	\caption{Evolution of net quark densities in the quark phase for $l=-b$.}
	\label{fig:figure3}
\end{figure}        

The evolution of the net lepton densities in the quark-gluon phase is shown in figure 
\ref{fig:figure2} for $l = -b$. At high temperatures there is no difference between the three lepton 
flavours. A negative net particle density means, that more antiparticles than particles exists. 
Below $\sim m_\tau/3 \approx 600$ MeV the tau leptons disappear, giving rise to an increase 
of net tau neutrino density, in order to keep $l_{\tau}$ constant. At the same time the number of 
positrons and positive muons increases in order to balance the positive charge that can no longer 
reside in the tau sector. To conserve flavour, this is compensated by a corresponding change in the neutrino sector.  

The net quark densities for this case are shown in figure \ref{fig:figure3}. The two different quark 
types $u$ and $d$ start at different points. If the temperature reaches the mass threshold of a quark species, it annihilates or decays and the remaining up- or down-like quarks increase their number densities to keep charge neutrality. For example the strange and down quarks react on the disappearance of the bottom quark. The figure shows, that considering only $u$, $d$, and $s$ quarks around the QCD transition at $T_{\rm QCD} \sim 200$ MeV is a reasonable approximation.

\begin{figure}     
        \centering
        \includegraphics[angle=270,width=0.8\textwidth]{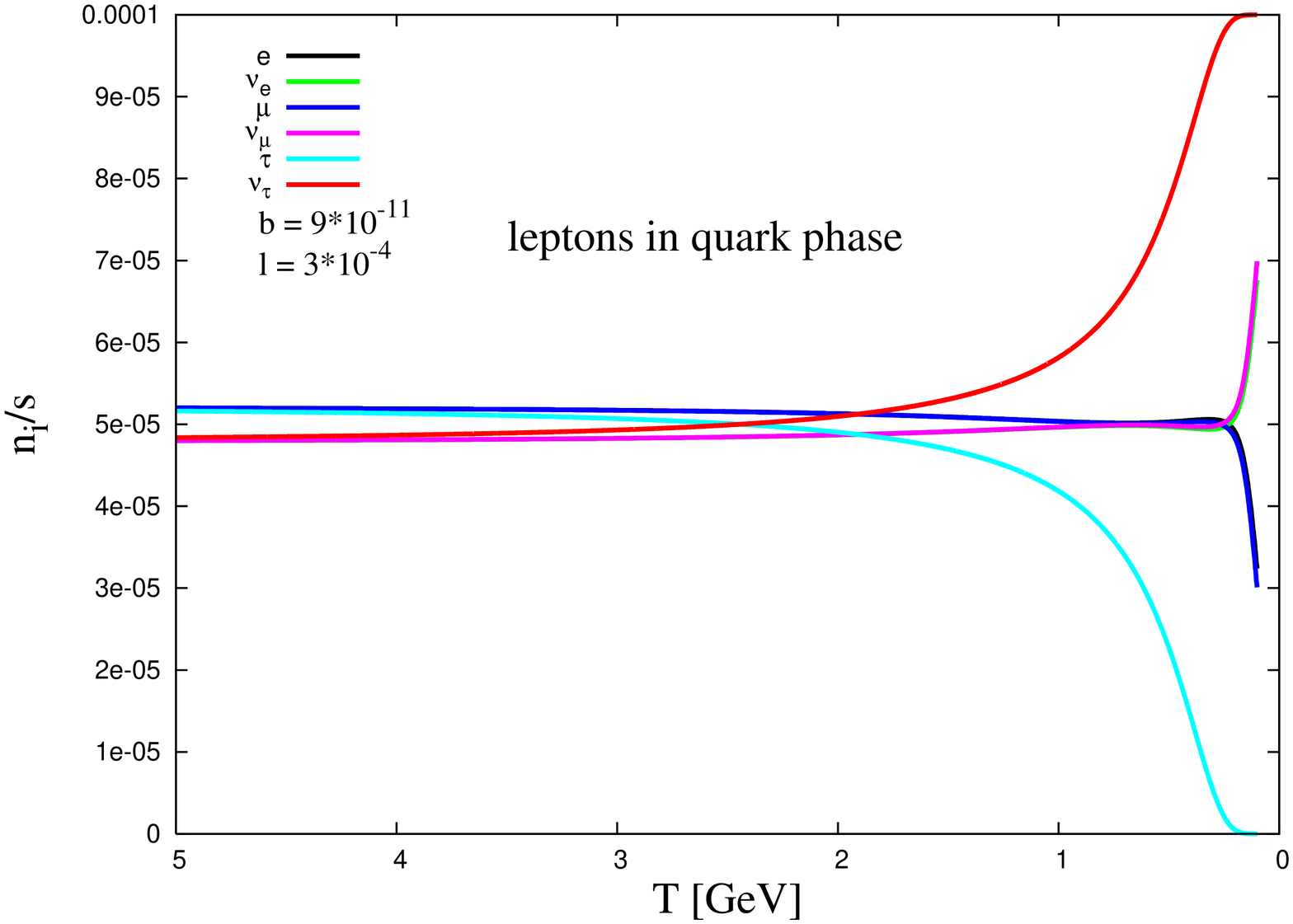}
        \caption{Evolution of net lepton densities in the quark phase for $l=3\times 10^{-4}$. The order 
        of magnitudes is very different from the case $l = - b$ and the sign is reversed.}
        \label{fig:figure4}
        \vspace{1cm}
        \includegraphics[angle=270,width=0.8\textwidth]{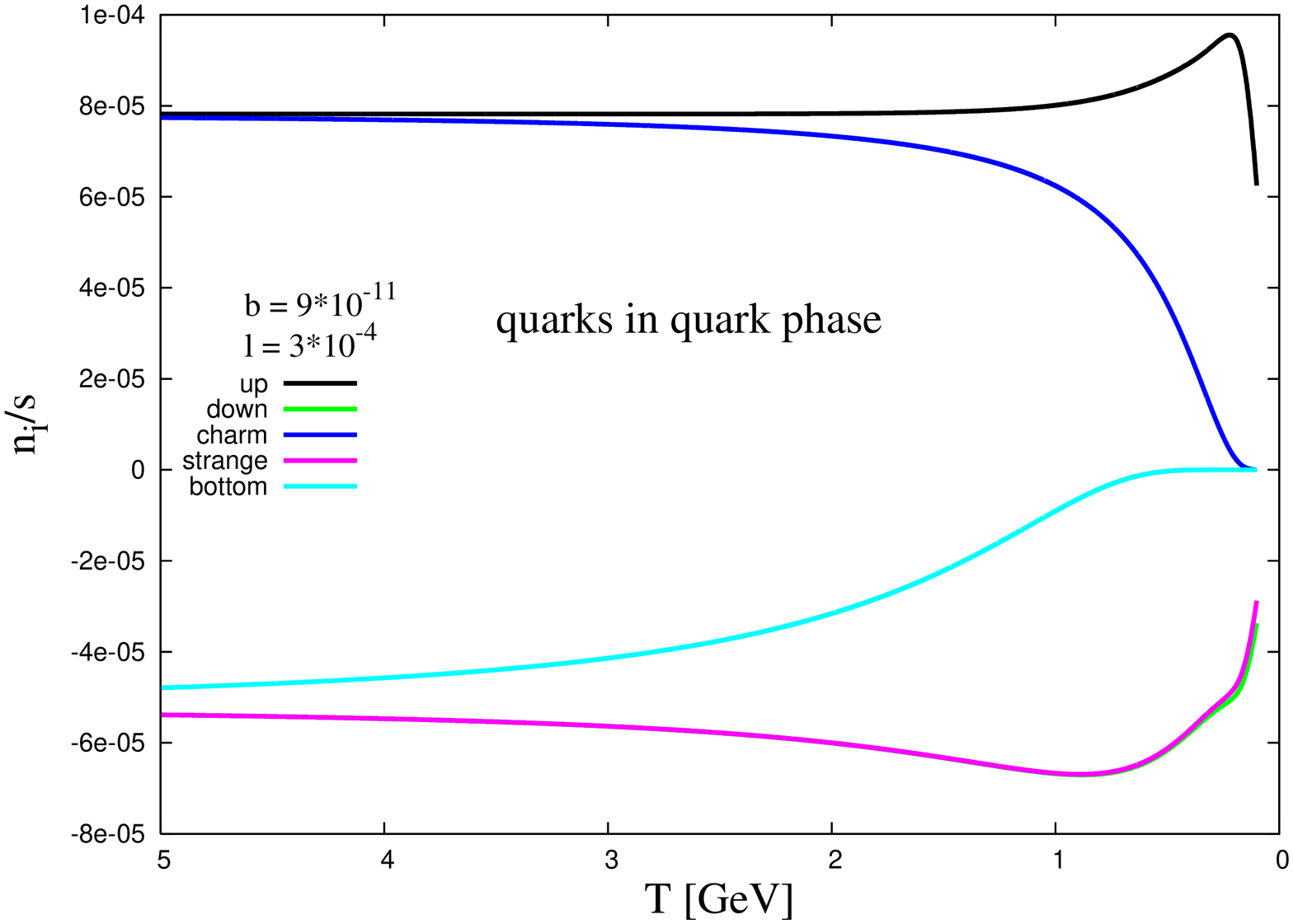}
        \caption{Evolution of net quark densities in the quark phase for $l=3 \times 10^{-4}$. 
        In comparison to the $l=-b$ scenario, we find a very different 
        behaviour of the net quark densities. Charge neutrality and the tiny baryon 
        asymmetry force the net densities of down-like quarks to negative values.}
        \label{fig:figure5}
        \end{figure}
  
Increasing the lepton asymmetry leads to drastic changes of net particle densities. In figures 
\ref{fig:figure4} and \ref{fig:figure5} their evolution in the quark-gluon phase are shown for 
$l=3 \times 10^{-4}$.
A positive lepton asymmetry leads to a dominance of negatively charged leptons. It comes without 
surprise that the net lepton densities react in a direct way, as they obviously depend on $l_f$. However, 
the quarks change their net densities by an comparable amount, as already indicated by our analytic estimates (\ref{muusimp}) and (\ref{mudsimp}). The net down-like quark densities become negative 
to react on the higher densities of the up-type quarks. The densities of the up types have to increase as an reaction of large lepton asymmetry to reach the charge neutrality. The different signs of the quark chemical potential for large $l$ are also indicated by (\ref{muusimp}) and (\ref{mudsimp}). Our numerical calculations confirm the analytic approach (\ref{muusimp}) to (\ref{muesimp}) and show that \textit{large lepton asymmetries influence the net quark densities strongly}.

\begin{figure}
        	\centering
        	\includegraphics[angle=270,width=0.8\textwidth]{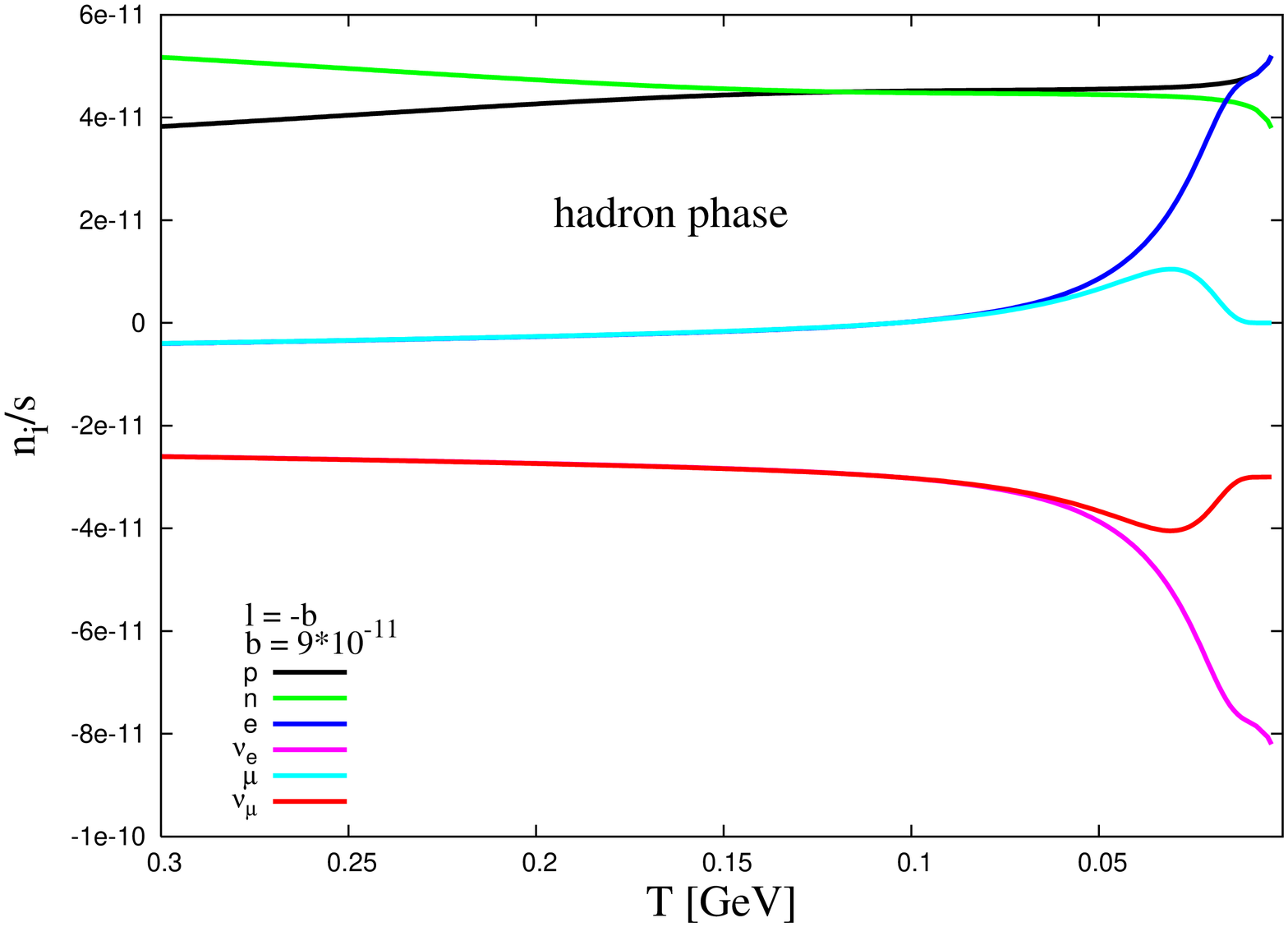}
         	\caption{Evolution of net particle densities in the hadron phase for $l= - b$. Mesons are included in the 											calculations but not shown.}
	\label{fig:figure6a}
         \vspace{1cm}
        \includegraphics[angle=270,width=0.8\textwidth]{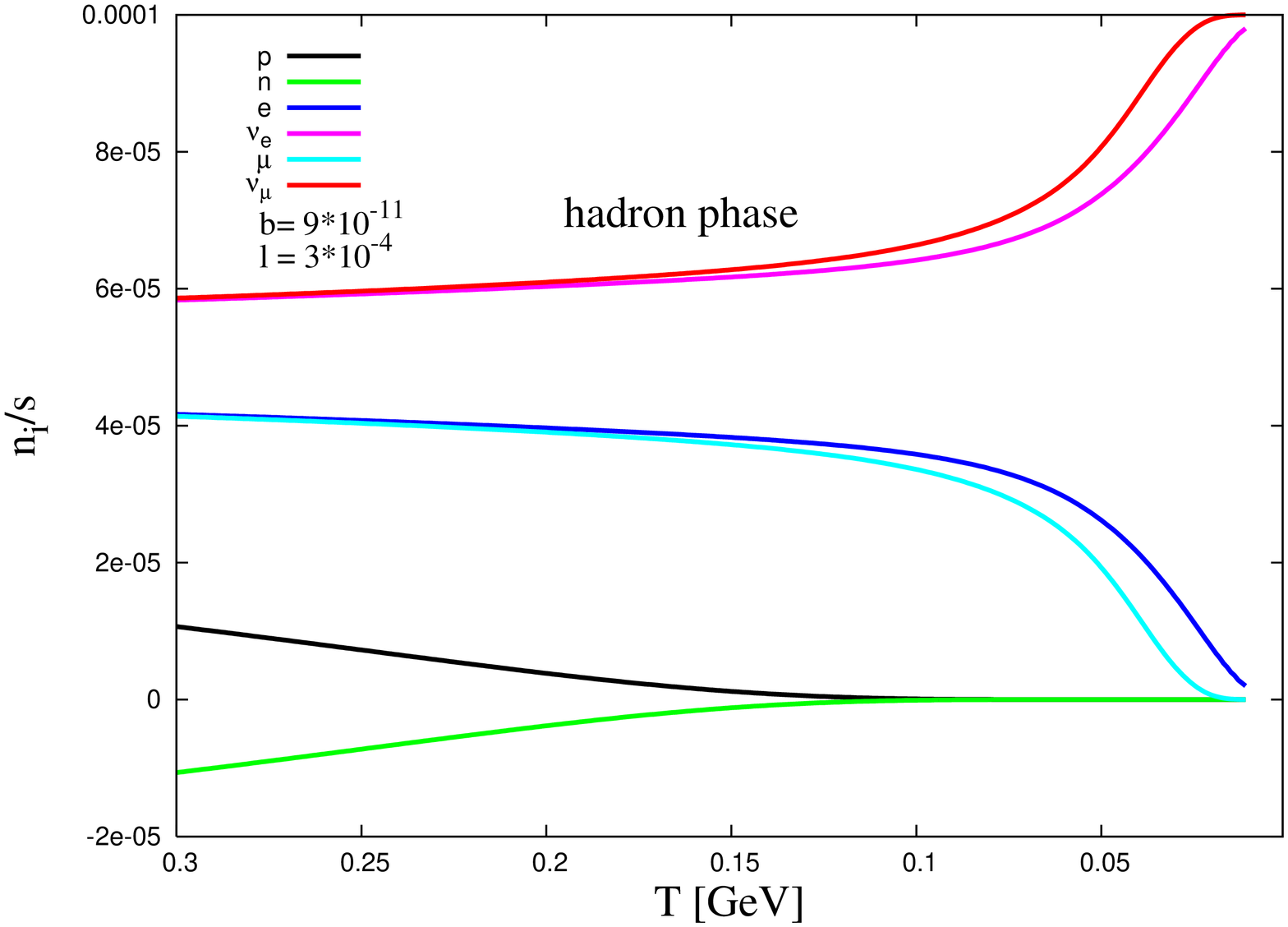}
        \caption{Evolution of net particle densities in the hadron phase for $l=3\times 10^{-4}$. 
        The influence of a larger $l$ on the net particle densities of the neutron and proton can 
        be seen clearly.}
        \label{fig:figure6b}
\end{figure}

Let us now turn to the hadron phase. The numerical solution for all relevant particles in the hadron phase are shown in figures (\ref{fig:figure6a}) and (\ref{fig:figure6b}). We consider the evolution form 
300 MeV down to 1 MeV. Below $T \sim 10$ MeV effects of neutrino oscillations become relevant, 
which are not included in this work. For these calculations we took all hadrons listed by the 
Particle Data Group up to the mass of $m_{K^{\ast}}=1414$ MeV into account.

The figures show the evolution of net particle densities for $l=-b$ and $l=3\times 10^{-4}$, respectively. 
Again it can be seen, that $l=-b$ leads to an antimatter dominated universe in the leptonic sector until 
pion disappearance (not shown in the figure). At the lowest temperatures considered, the electron density is strongly linked to the baryon number via the charge neutrality of the Universe. The inclusion of light mesons like pions and kaons is also important to understand the split between 
proton and neutron densities at temperatures close to the QCD transition. This is another effect
that is not obvious from the very beginning. As pions are effectively massless, they can carry negative 
charge density in order to compensate for the positive charge density in protons and positrons. 
The influence of larger $l$ on the net particle densities of the neutron and proton can be clearly 
seen, as the evolution differs in the high temperature regime. For low temperatures their evolution is independent of $l$, as indicated by (\ref{mupHana}). The effect of large $l$ on $\mu_p$ and 
$\mu_n$ for small $T$ is negligibly.
 
Our numerical calculation fits the analytic approaches for both phases well. There is clearly an 
influence of large $l$ on quarks and hadrons at the QCD transition.

\section{Chemical potentials}

Each conserved quantum number can be associated with a chemical potential. The particles contribution to the free energy  can then be described as 
\begin{eqnarray}
\fl \mu_Qn_Q + \mu_Bn_B + \sum_f\mu_{L_f}n_{L_f} 
& 
\stackrel{T>T_{\rm QCD}}{=} \sum_q\mu_qn_q + \sum_l \mu_{l}n_{l} + \sum_{g}\mu_{g}n_{g} 
\nonumber \\
&
\stackrel{T<T_{\rm QCD}}{=} \sum_b \mu_b n_b + \sum_{m}\mu_{m}n_{m} + \sum_l\mu_{l}n_{l}, \nonumber
\end{eqnarray}
with Q denoting charge, B baryon number, and $L_f$ lepton flavour number on the l.h.s. and
q quarks, l leptons, g massive gauge bosons (the gluon and photon chemical potentials vanish), b baryons and m mesons.

The particle fluid in the quark phase contains all types of quarks and leptons. 
The massive gauge bosons are way too heavy to play a crucial role at the QCD epoch, 
nevertheless we keep them in our model. So we get: 
\begin{eqnarray}
&\mu_Q(\frac{2}{3}n_u-\frac{1}{3}n_d+\frac{2}{3}n_c-\frac{1}{3}n_s+\frac{2}{3}n_t-\frac{1}{3}n_b-n_e-n_{\mu}-n_{\tau} + n_W)\nonumber \\ &
+\mu_B\left(\sum_q\frac{1}{3}n_q \right)+ \sum_f \mu_{L_f} (n_f + n_{\nu_f}) \nonumber \\
 =&\sum_q\mu_qn_q+\sum_l\mu_{l}n_{l}+\mu_Wn_W. \nonumber
\end{eqnarray}
The Z-boson does not show up, as at high energies its chemical potential vanishes (like the one of photons and gluons) and there are no Z-bosons at low temperatures. For the W we pick the convention that $n_W$ counts $W^+$ with a plus sign. 
A comparison of the coefficients in front of the net particle densities gives:
\numparts
\begin{eqnarray}
\mu_u=\mu_c=\mu_t=\frac{2}{3}\mu_Q+\frac{1}{3}\mu_B, \\
\mu_d=\mu_s=\mu_b=-\frac{1}{3}\mu_Q+\frac{1}{3}\mu_B, \\
\mu_f=-\mu_Q+\mu_{L_f}, \\
\mu_{\nu_f}=\mu_{L_f}, \\
\mu_W=\mu_Q.
\end{eqnarray}
\endnumparts
Consequently, the chemical potentials of the globally conserved quantities in the quark phase are 
\numparts
\begin{eqnarray}
\label{muBhigh}\mu_B(T>T_{\rm QCD}) = \mu_u+2\mu_d, \\
\label{muQhigh}\mu_Q(T>T_{\rm QCD}) = \mu_u-\mu_d, \\
\label{mulfhigh}\mu_{L_f}(T>T_{\rm QCD}) = \mu_{\nu_f}.
\end{eqnarray}
\endnumparts

The conserved quantum numbers in the hadron phase are described by
\begin{eqnarray}
\fl \mu_{Q}(n_p+n_{\pi}+n_K+\dots-n_e-n_{\mu}) +
\mu_B(n_p+n_n) + \sum_f\mu_{L_f}(n_f+n_{\nu_f}) \nonumber \\
\fl = \sum_b\mu_b n_b +\sum_{m}\mu_{m}n_{m}+\sum_l\mu_ln_l. \nonumber
\end{eqnarray}
As above, a comparison of the coefficients leads to
\numparts
\begin{eqnarray}
\label{muBlow}\mu_B(T<T_{QCD})=\mu_n, \\
\label{muQlow}\mu_Q(T<T_{QCD})=\mu_p-\mu_n=\mu_{\pi}, \\
\label{mulflow}\mu_{L_f}(T<T_{QCD})=\mu_{\nu_f}.
\end{eqnarray}
\endnumparts

\subsection{Baryochemical potential}

The baryochemical potential $\mu_B$ is the energy needed to add a baryon to a thermalised 
state at fixed volume and entropy. In general, it depends linearly on $b$ and $l$. 

In the quark phase, taking only three quarks (u,d,s), electrons and muons into account (and 
neglecting their masses), we insert (\ref{muusimp}) and (\ref{mudsimp}) 
into (\ref{muBhigh}) to find $\mu_B(T > T_{\rm QCD}) =  3 b s(T)/ T^2$. Thus, it seems that the baryochemical potential would just depend on the baryon asymmetry, as one could naively expect.
However, if we consider a slightly higher temperature, such that we have to include charm and tau, 
we find from (\ref{muu1GeV}) and (\ref{mud1GeV}) that 
\begin{equation} 
\label{muBh}
\mu_B(T \gg T_{\rm QCD}) =  \left(\frac{39}{4} b  -  l\right) \frac{s(T)}{4 T^2}. 
\end{equation}
In general, $\mu_B = \mu_B(b,l)$. 
Thus, for $|l| \gg b$, the order of magnitude of the baryochemical potential is set by the lepton asymmetry, and not as naively expected by the baryon asymmetry.  

This can be understood by the following consideration: increasing the lepton flavour asymmetries, the corresponding chemical potentials are forced to react, see (\ref{leAna}) to (\ref{ltauAna}). On the other hand, electric charge and baryon number must not be changed.  A large net charge in the leptons 
must be compensated by a large and opposite net charge carried by quarks, in a way that the 
baryon asymmetry remains tiny. That makes it very difficult to add a baryon without disturbing that fine balance, and that is why the baryochemical potential (\ref{muBhigh}) grows as large as the 
lepton asymmetry.

\begin{figure}
	\centering
		\includegraphics[angle=270, width=0.80\textwidth]{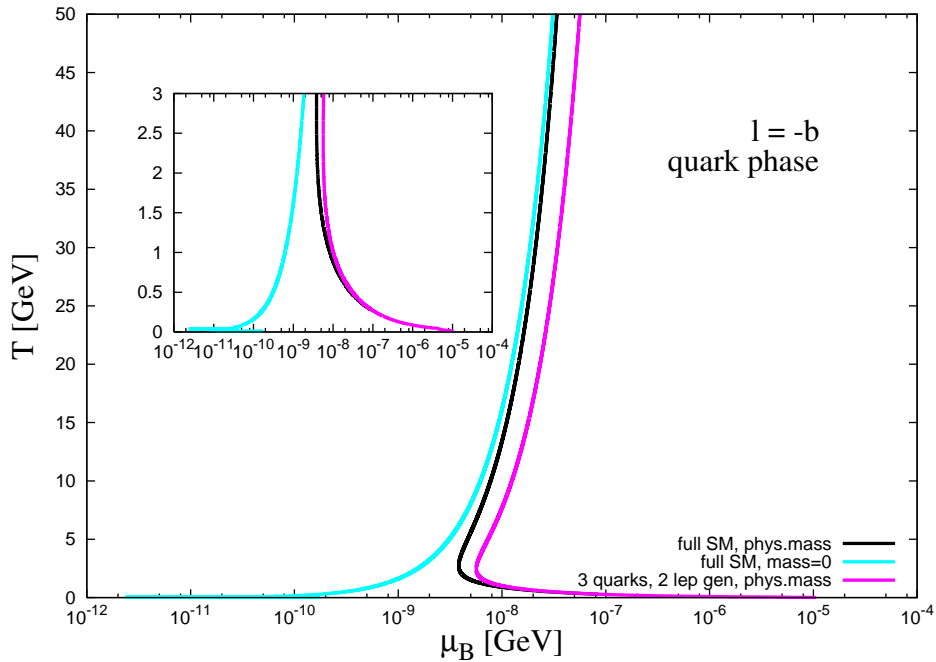}
	\caption{Trajectory of the quark phase in the $\mu_B-T$ plane for the $l=-b$ scenario. We compare several approximations to the exact result (black line). At high temperatures all particles can be assumed to be massless (blue line), while below $\sim 5$ GeV mass thresholds are important. Below 1 GeV the universe is  well described by three quark (up, down, strange) and two lepton (electron, muon) flavours only (magenta line). }
	\label{fig:figure8}
\end{figure}

Neglecting the physical masses of the particles would have a strong effect on $\mu_B$, as shown in figure \ref{fig:figure8}. While the simple analytic approach with $m=0$ gives $\mu_B \propto T$ (\ref{muBh}), the numerical results with physical quark masses show in the low temperature regime of the quark phase a different behaviour. The non-relativistic particles give rise to an increase of $\mu_B$ below $T\sim 3$ GeV, i.e. charm, bottom quarks and taus play an important role. However, neglecting the heavy quarks, but taking the physical masses of the strange quark and the muon into account, gives a reasonable approximation at the QCD scale (see figure \ref{fig:figure8}). 

In the hadron phase at low temperatures, but above temperatures of a few MeV, the difference in the masses and chemical potentials of the proton and neutron is negligible and one can assume $m = m_p \approx  m_n$ and $\mu_B = \mu_n \approx \mu_p$, as given by (\ref{lowTmup}) after pions and 
muons disappeared from the thermal bath. Thus, the baryochemical potential becomes
\begin{equation}
\label{muBsmall}  
\mu_B(T< m_\pi/3) \approx m - T \ln\left[\frac{c(T)}{2 b s(T)} \right].
\end{equation}
In the low temperature regime, $\mu_B$ is linear in $T$ and approaches the nucleon mass $m$.

Let us now turn to a more detailed numerical study. We evaluate $\mu_B(T)$ numerically in the quark and hadron phases, i.e.~expressions (\ref{muBhigh}) and (\ref{muBlow}) respectively, for various 
values of $l$. 

\begin{figure}
	\centering
	\includegraphics[angle=270,width=0.80\textwidth]{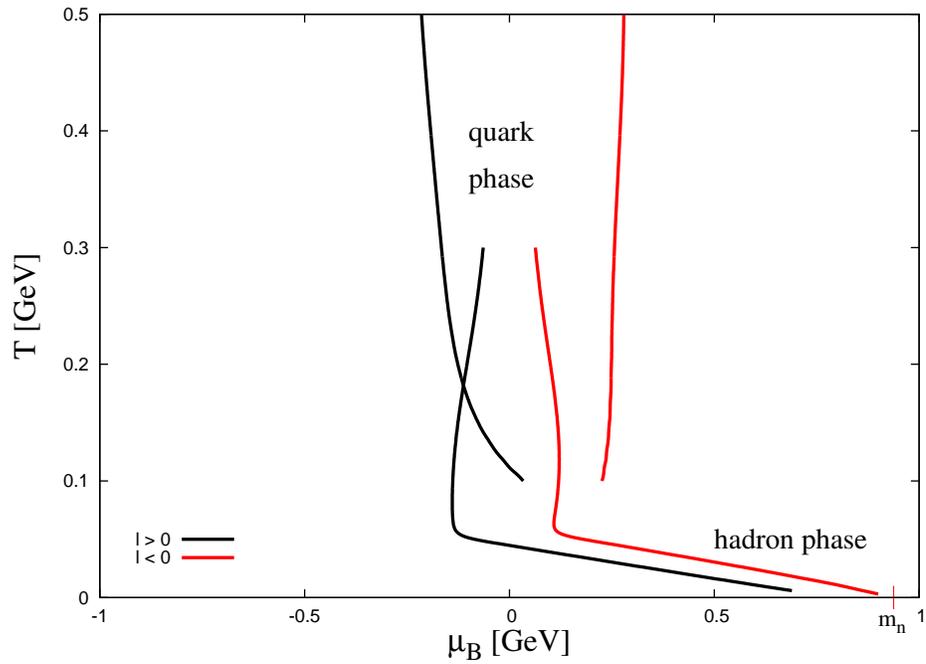}
	\caption{Evolution of the baryochemical potential in the quark and hadron phase 
	for large lepton asymmetry $l= \pm 3 \times 10^{-2}$. 
	Lepton asymmetries of different signs lead to qualitatively different trajectories. 
	While the trajectories of the quark and hadron phase cross for positive $l$, they miss each 
	other for negative $l$ in the expected temperature region at $\sim 200$ MeV.}
	\label{fig:figure9}
\end{figure}
\begin{figure}
	\centering
	\includegraphics[angle=270,width=0.80\textwidth]{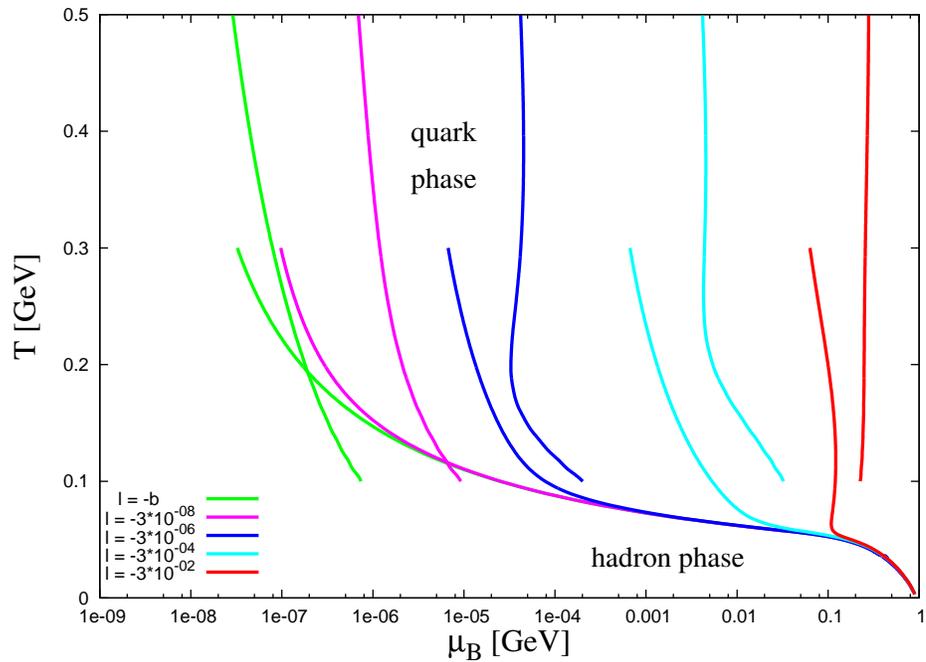}
	\caption{Evolution of the baryochemical potential for negative lepton asymmetries. 
	Above the pion threshold, the trajectories of both phases depend strongly on the 
	lepton asymmetry. For lower temperatures the lepton asymmetry is negligible.}
	\label{fig:figure10}
\end{figure}

If there is no baryo- or leptogenesis after sphaleron processes cease, the standard model predicts a negative value for $l$. The influence of the sign of $l$ on the baryochemical potential is displayed in figure \ref{fig:figure9}. In order to illustrate the effect of the sign, we choose a rather ``large'' value for the lepton asymmetry $l = \pm 3 \times 10^{-2}$. 
For negative $l$, $\mu_B$ is positive for all temperatures. In contrast, for positive values of $l$, 
$\mu_B$ can take negative values. For the high temperature regime, this is easily seen from
(\ref{muBh}). In the hadron phase, $\mu_B$ approaches the nucleon mass at low temperatures, independently of the value of $l$. 
While a positive $l$ always allows for a point of coexistence of the quark and hadron phase at a temperature of $\sim 200$ MeV, the trajectories of the $l<0$ case would meet at a much lower temperature.

Figure \ref{fig:figure10} shows the case of negative lepton asymmetry for different values of $l$. 
The influence of the lepton flavour asymmetry on the baryochemical potential is evident. 
The point of coexistence moves to larger baryochemical potentials and lower temperatures as 
$|l|$ increases (not all crossings are shown in the figure). 
Shortly after the transition to the hadron phase, the trajectories are still heavily affected by the
lepton asymmetry,  as pions and other mesons can carry a net charge density. 
For temperatures below the pion threshold, $T < m_{\pi}/3 \approx 50$MeV, the hadron phase 
follows a unique trajectory, as predicted by (\ref{muBsmall}). 

\subsection{Charge chemical potential}

We associate with the globally conserved and vanishing charge of the Universe the charge 
chemical potential $\mu_Q$, which is the energy needed to add a charge unit to a 
thermalised state at fixed volume and entropy.

For $T>T_{\rm QCD}$ it depends, like the baryochemical potential, on $\mu_u$ and $\mu_d$. But the 
effect of a large lepton asymmetry on the evolution is bigger because of the missing factor of two for
$\mu_d$. From the approximations (\ref{muusimp}) and (\ref{mudsimp}), for the case of three quark and two lepton flavours, $\mu_Q = l s(T)/2T^2$.
Taking, as above,  the charm and tau into 
account and using (\ref{muu1GeV}) and (\ref{mud1GeV}) we find:
\begin{equation}
\mu_Q (T\gg T_{\rm QCD}) = \left( -\frac{3}{4}b + l\right)\frac{s(T)}{2T^2}.
\label{muQTbig}
\end{equation}
The charge chemical potential is now depending on $b$ too, and does not even vanish in the case 
$b = l$, as one could naively expect. Taking the particle masses into account has the same effect, 
namely $\mu_Q = \mu_Q(b,l)$ in general.  

After the QCD transition, a charge unit can be added to the Universe in the form of either a charged 
pion, muon or electron. At $T\approx m_{\pi}/3 \approx 50$ MeV, pions annihilate and the charge neutrality has to be ensured by protons and electrons. 
Using (\ref{mupmun}), (\ref{mupHana}) and (\ref{muQlow}) one gets 
\begin{equation} 
\mu_Q(T < T_{\rm QCD}) = \left( -9 b + 4l\right) \frac{s(T)}{2T^2 + \frac{9c(T)}{T}\rm{exp}[-m/T]}. 
\end{equation} 

\begin{figure}
	\centering
	\includegraphics[angle=270,width=0.80\textwidth]{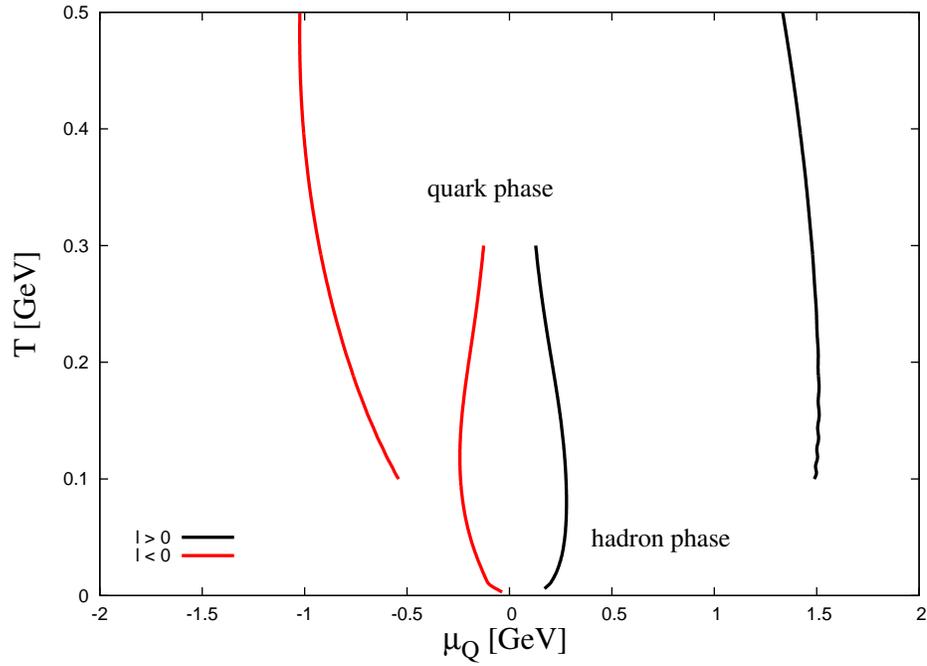}
	\caption{Evolution of the charge chemical potential for $l = \pm 3 \times 10^{-2}$. The trajectories of the quark and hadron phase miss each other for both signs of $l$.}
	\label{fig:figure11}
\end{figure}

\begin{figure}
	\centering
	\includegraphics[angle=270,width=0.80\textwidth]{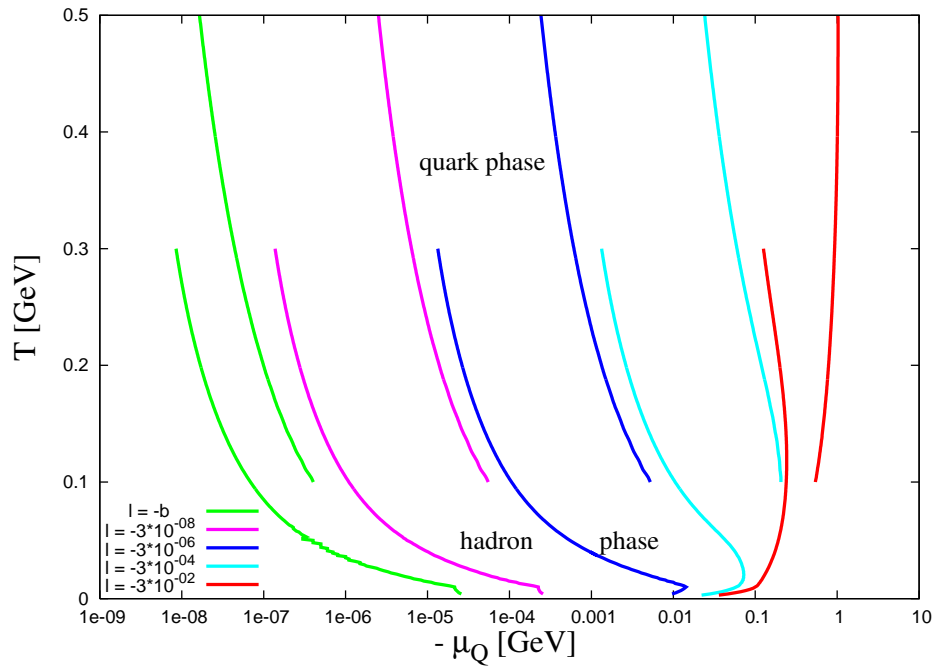}
	\caption{Evolution of a negative charge chemical potential for negative values of $l$.}
	\label{fig:figure12}
\end{figure}

The dependence of $\mu_Q$ on the sign of $l$ is shown in figure \ref{fig:figure11}. 
In the quark phase it behaves like $\mu_Q\propto l/T^2$, so that the sign of $l$ fixes the sign 
of $\mu_Q$. Figure \ref{fig:figure12} shows the dependence of $\mu_Q$ for negative $l$. A negative 
value of $l$ implies a negative charge chemical potential.
It is interesting that there does not exist any point of coexistence of phases for all the studied cases. 


\subsection{Leptochemical potentials}

With each lepton flavour a chemical potential is associated. It is the energy needed to add a lepton of a certain flavour to a thermalised state at fixed volume and entropy. The cheapest way to add a 
lepton of some flavour is to add a corresponding neutrino, see (\ref{mulfhigh}) and (\ref{mulflow}). 
While for the case of three light quarks and two charged leptons at temperatures just above the QCD transition, $\mu_{L_e} = \mu_{L_\mu} = \mu_{L_\tau}/2 = ls(T)/T^2$ from (\ref{muesimp}) and 
(\ref{ltauAna}), including the charm and tau 
at higher temperatures leads to (\ref{munu1GeV}) 
\begin{equation}
\mu_{L_f}(T \gg T_{\rm QCD}) =  \left(l-\frac{1}{4}b \right)\frac{s(T)}{T^2}
\end{equation}
for all three lepton flavours. Again, taking the finite masses of quarks and charged leptons into account 
shows that $\mu_{L_f} = \mu_{L_f}(b,l)$. 

After the annihilation of the tau and muon, the flavour asymmetries survives in the neutrino sector. Only the asymmetry in the electron flavour can be divided into electron and electron neutrino asymmetry. This changes with the onset of the neutrino oscillations at $T \sim 10$MeV. The flavours oscillate and the three neutrino sectors become linked via 
$\nu_{e}\rightleftharpoons \nu_{\mu}\rightleftharpoons \nu_{\tau}$ and lepton flavour is no longer  
a conserved quantum number. However, the total lepton number still is (if neutrinos are not their own
anti-particles, i.e.~are not Majorana particles). 
 
\begin{figure}
	\centering
	\includegraphics[angle=270,width=0.80\textwidth]{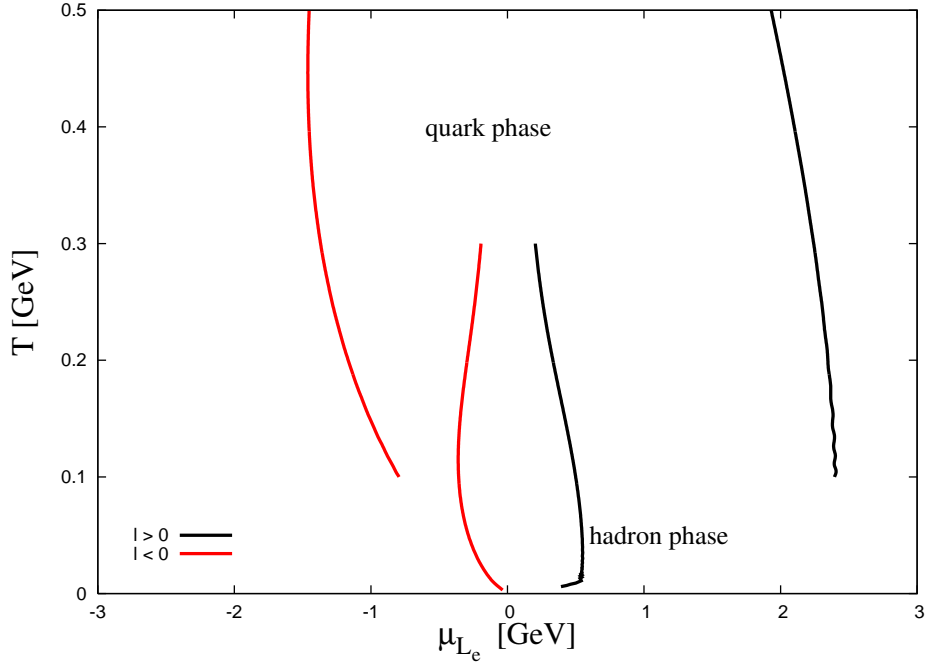}
	\caption{Evolution of the leptochemical potential associated with the first lepton generation
	for $l = \pm 3 \times 10^{-2}$.}
\label{fig:figure13}
\end{figure}
\begin{figure}
	\centering
	\includegraphics[angle=270,width=0.80\textwidth]{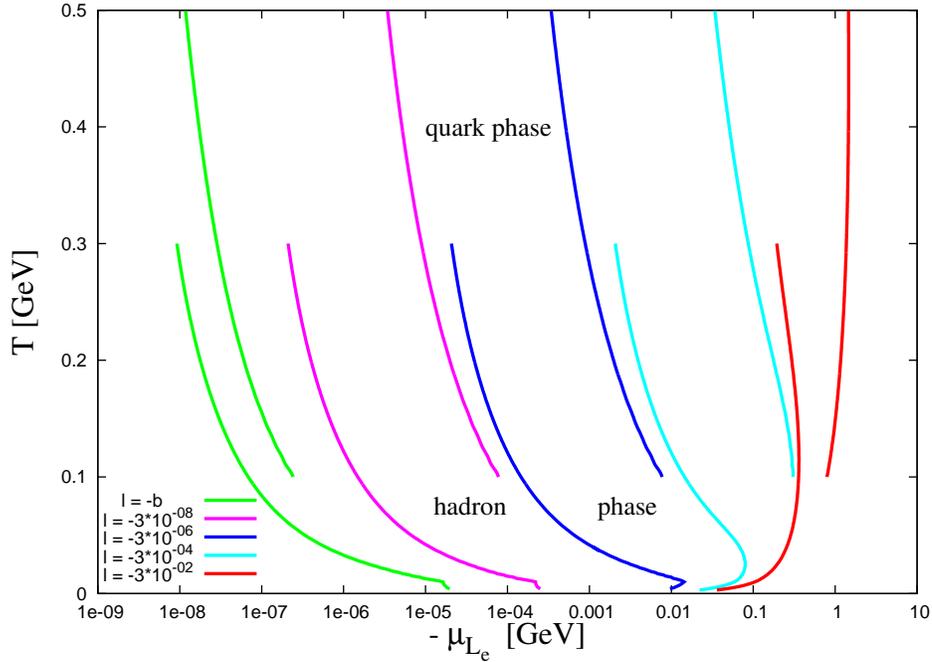}
	\caption{Evolution of a negative leptochemical potential associated with the first lepton generation
	for negative $l$.}
\label{fig:figure14}
\end{figure}

\section{Consequences for the cosmic QCD transition} 
 
Our findings question some of the established results about the dynamics of the cosmic QCD transition. It was concluded from lattice QCD simulations, that the cosmic QCD transition is a crossover \cite{Aoki:2006we}. These simulations included three dynamical quarks and implicitly assumed $\mu_B=\mu_Q=0$. However, $\mu_B$ and $\mu_Q$ vanish only in the case of $b=l=0$. There also exist first lattice studies for tiny baryon and charge chemical potentials \cite{Cheng:2008zh}. It seems to us, that tiny, non-vanishing $b$ and $l$ should not change this story. While $b$ is certainly tiny in nature, this is not necessarily true for $l$. A large lepton asymmetry results in a large $\mu_B={\cal O}(l)T$ and $\mu_Q={\cal O}(l)T$ at $T\simeq T_c$. Thus, we have to revisit the question of the order of the cosmic QCD transition.

At any thermodynamic transition, the chemical potentials of conserved quantum numbers must evolve continuously.
For the scenario $l \sim -b$ we found for the baryochemical potential a crossing of the quark and hadron phase trajectories at $T \approx 190$ MeV (figure \ref{fig:figure10}). This seems to be consistent with the transition temperature found by lattice simulations. Increasing the lepton asymmetry seems to lead to a decrease of the critical temperature. The opposite effect occurs at the electroweak transition, shown in \cite{Gynther:2003za}. In contrast to the trajectory of the baryochemical potential, the trajectories for the charge and leptochemical potentials do not cross at $T\sim 200$ MeV. Taken seriously, this would require a jump from the quark- to the hadron trajectory, which would lead to a non-thermal transition. However, as we have neglected interactions in our analysis, the trajectories are modified close to $T\sim \Lambda_{QCD}$ and our findings just tell us, that the situation of $\mu_Q \neq 0$ and $\mu_{L_f} \neq 0$ has to be studied carefully, i.e. by means of lattice simulations.   

The sign of a possible lepton asymmetry is important for the dynamics of the transition, see figures \ref{fig:figure9}, \ref{fig:figure11} and \ref{fig:figure13}. A negative lepton asymmetry leads to negative charge- and leptochemical potential and a positive $\mu_B$. If the asymmetry is positive, the signs of the chemical potentials change. For large $|l|$ or for positive $l$, the trajectories of the quark and hadron phases do not cross in the $\mu_B-T$ plane (at least not at $T> 100$ MeV). 
Lepton asymmetries of the order of $10^{-2}$ might put the early Universe in the vicinity of the critical 
end point $(T,\mu_B)=(T_{\rm e},\mu_{\rm e})$, see figure 1.  

To go into more details is beyond the scope of this work, but it is clear, that the phase diagram for 
the cosmic phase transition lives in 5 dimensions for a charge neutral universe (3 lepton flavour asymmetries, the baryon asymmetry and temperature). So far just a two dimensional slice of it has been studied in greater detail.   

\section{Conclusions}

We know very little about the actual lepton asymmetry of the Universe. A large lepton 
asymmetry $|l| \gg b$ is compatible with constraints form primordial nucleosynthesis and the CMB, 
if $|l| < 0.02$ \cite{Simha:2008mt,Popa:2008tb}.  We have shown that such an asymmetry may significantly affect the dynamics of the cosmic QCD transition. The cosmic trajectory in the $\mu_B$-$T$ plane depends on 
the lepton flavour asymmetries $l_f$, besides the baryon asymmetry $b$. We explicitly discussed the dependence of the baryochemical potential on the lepton asymmetry, $\mu_B=\mu_B(T,b,l)$. While in systems, where only strong interactions matter (relativistic heavy-ion collisions, lattice QCD) a large $\mu_B$ implies large baryon densities, the early universe can support large $\mu_B$ via a large lepton asymmetry.  

Depending on the, yet unknown, structure of the QCD phase diagram and especially on the 
position of a hypothetical critical end point, a large lepton asymmetry might result in a first-order 
QCD transition in the early Universe. This seems possible for $|l| \simeq 0.02$. Many formerly 
suggested cosmological consequences of a first-order QCD transition would be possible (formation of relics, effects on nucleosynthesis, generation of magnetic fields, generation of gravitational waves, etc. 
\cite{Schwarz:2003du}).

The conclusions from lattice QCD simulations at $\mu_Q = 0$ and $\mu_B \ll T$ have been used to conclude that the cosmic QCD transition is a crossover \cite{Aoki:2006we}. We have shown, 
that the cosmic trajectory has $\mu_Q \neq 0$. In the case of efficient sphaleron processes, $\mu_Q$ is tiny. In the case of a large lepton asymmetry, $\mu_Q$ can be large and it is unclear if the conclusion that the cosmic QCD transition is a crossover remains true. For $|l|>10^{-4}$, the charge chemical potential differs between the quark and the hadron phase by more than 100 MeV at temperatures around the QCD scale (see figure \ref{fig:figure12}).But note, that we neglected all interactions in our analysis. 

A detailed understanding of the consequences of $|l| \gg b$ on the QCD transition could allow us to rule out or find evidences for leptogenesis scenarios that lead to a large lepton asymmetry. This is 
of interest to all scenarios in which the lepton number is changed after the sphaleron processes stop to be efficient (late leptogenesis e.g. \cite{Hambye:2001eu}). 

In this work we assumed that all globally conserved quantum numbers are also conserved locally and we have put the focus on the case of equal lepton flavour asymmetries. Dropping the latter, obviously 
leads to a somewhat richer phenomenology, which is beyond the scope of this work. 
While the local conservation of electric charge and baryon number are probably excellent approximations, the local conservation of lepton number is not correct, if distance scales below the 
neutrino mean free path are considered. Consequently, as soon as inhomogeneities become important (and they will in the case of a first-order phase transition, as cold spots are more likely to nucleate bubbles of the new phase \cite{Ignatius:2000cz}) it would be more realistic to describe an inhomogeneous universe with regions of different $l_f$. These different regions would be equilibrated via neutrinos, so that possible bubbles of different $l_f$ at the QCD phase transition would have a minimal radius, given by the mean free path, 
$d_{\nu-{\rm mfp}} \simeq 1$ m \cite{Ignatius:2000cz}. This has to be compared to the size of the Universe at this time, $d_H \simeq 10$ km \cite{Schwarz:2003du}. Although giving rise to small effects only, precise measurements of the abundance of primordial light elements might be sensitive to inhomogeneities produced during the QCD transition, as they might lead to inhomogeneous nucleosynthesis. 

Some consequences of a large lepton asymmetry on the physics of the early Universe in the QCD epoch have been overlooked so far. This calls for a reinvestigation of many of the suggested	
effects of the cosmic QCD transition and might allow us to improve the limits on the universal lepton flavour asymmetries (before neutrino oscillations start). 
  
\ack %acknowledgement

We thank Stefan Froehlich, Thomas Hambye, Olaf Kaczmarek, Frithjof Karsch, Edwin Laerman, Mikko Laine and Christian Schmidt for comments, discussions and help with the numerical calculations. We acknowledge financial support from the Deutsche Forschungsgemeinschaft (GRK 881). M.S.~is supported by the 
Friedrich-Ebert-Stiftung.

\end{document}